\documentclass[english,11pt,condensed]{article}
\usepackage[T1]{fontenc}
\usepackage[latin9]{inputenc}
\usepackage{color}
\usepackage{babel}
\usepackage{ragged2e}
\usepackage{cite}
\usepackage{geometry}
\usepackage{graphicx}
\usepackage[unicode=true,pdfusetitle,
 bookmarks=true,bookmarksnumbered=false,bookmarksopen=false,
 breaklinks=false,pdfborder={0 0 1},backref=false,colorlinks=true]
 {hyperref}

\makeatletter
\newcommand{\lyxaddress}[1]{
\par {\raggedright #1
\vspace{0.4cm}
\noindent\par}
}
\date{}

\makeatother

\begin{document}
\title{ {\normalsize ~ \hfill CERN-TH-2016-009 } \\ ~ \vspace{2.0 cm} \\ Wish-list for Run II top quark measurements (for BSM)}

\author{Roberto Franceschini}

\maketitle

\lyxaddress{\begin{center}
\vspace{-0.8 cm}Theoretical Physics Department, CERN, Geneva, Switzerland
\par\end{center}}

\begin{abstract}
In this contribution I will highlight the new challenges for top quark physics at LHC Run II, focusing in particular on the interplay between precision studies on the top quark and searches for new physics. A new strategy to search for subtle scenarios of new physics is envisaged. The ability of compute and measure very accurately top quark properties such as its production rate, decay properties, {\it e.g} rates and  distributions, and variables sensitive to the top quark mass is put at the center of this strategy to probe new physics.\end{abstract}

\section{Introduction}

The Large Hadron Collider is known for being a ``top quark factory'', with a cross-section for top quark pair production nearing one $nb$, meaning millions of top quarks produced each year of running. Such a large rate for production of top quarks opens the way to {\it precision studies} of top quarks properties that were simply unattainable at previous machines. With such large top quark sample already recorded, or to be recorded in the next few years during Run II, statistical uncertainties are on their way to become less and less a limitation to the ultimate precision for experimental measurements. This prospect has stimulated a very intense effort for improving theory calculations to a level of precision that can match the foreseen experimental one. Presently it is possible to compute top quark production~\cite{Czakon:2015owf} and decay~\cite{Gao:2013wk,Brucherseifer:2013mg} at NNLO in the strong coupling constant; huge progress has been made in matching NLO computations and parton showers \cite{Campbell:2014fr} and computation of off-shell effects \cite{Heinrich:2014tg,Bevilacqua:2011fv,Denner:2012ul}; furthermore, automated NLO computations \cite{Alwall:2014bq} can be used to produce distributions for any observable the experiments want to measure.

A similarly flourishing activity has taken place in many other areas of precision calculations for hadron collisions and the results of Run I of the Large Hadron Collider have yield a wonderful agreement between precision calculations of Standard Model processes and measurements by the Large Hadron Collider experiments, as witnessed by the summary of the Standard Model (sub)groups at this conference and their updates~\cite{ATLASsmSummary}.

The great advances of precision measurements and precision calculations for hadron colliders is not only an enormous achievement {\it per se}, it also enables new strategies for the search of new physics at hadron colliders. In fact, hadron colliders are traditionally considered ``discovery machines'' for their ability to reach the highest energies and at the same time being able to scan for new phenomena on a vast range of energy scales. These features arise from the possibility to easily accelerate protons and effectively collide their constituents (quarks and gluons) with a large rate and a broad range of center of mass energies. At the same time the structure of the colliding protons poses a serious challenge to our ability to calculate, for instance because of possible interactions between initial states and final states of the collisions, and, in addition to this conceptual obstacle, perturbation theory of strong interactions converges rather slowly, hence several perturbative orders are sometimes needed to go seriously beyond an order of magnitude estimate of certain quantities. 

The latest achievements in precision calculations have brought our understanding of hadronic collisions to a level that, as I will argue in the following, they can presently be used to spot new physics in subtle deviations from the predictions of precision calculations for observables that are sensitive to new physics. As a matter of fact, this use of precision calculations at hadron colliders opens the way for investigations of new physics with a mindset that is rather more typical of leptonic machines.

\section{Subtle New Physics At Run II}
Despite the reduced beam energy compared to design at which the Large Hadron Collider has run in the past few years, already at this early stage of the life of the Large Hadron Collider a large amount of new physics scenarios have been tested. With supersymmetry being a leading candidate for the model of physics of the TeV scale, the impact of the Large Hadron Collider on new physics searches can be effectively summarized saying that colored superparticles are bound to be heavier than 1 TeV in most supersymmetric scenarios~\cite{Dine:2015qc,Craig:2013yu,Feng:2013ys}.  

These tight bounds  hold for a large class of models, however several possible ways out of this bound have been suggested. The details of the models that allow to alleviate the bounds that new physics scenarios receive from searches at the Large Hadron Collider cannot fit these pages, a sampler of options can be found in the literature \cite{Serra:2015ai,Anandakrishnan:2015tg,Kribs:2012qy,Papucci:2011rc,Franceschini:2015xy,Das:2012ya}. Nevertheless a unifying features of these models can be easily identified: all these models, in order to relax constraints from experimental searches, need to predict less apparent signatures of new physics, on which the experiments have a hard time to put bounds on. The increased difficulty to put bounds on these scenario can originate from {\it i)} lower production cross-section for the new physics states, which in turn can originate from having fewer new physics particles accessible at the Large Hadron Collider, or fewer available mediators for the production mechanisms ({\it e.g.} suppression of strong interaction production, removal of certain associated production mechanisms) {\it ii)} less visible signatures, which  feature  fewer leptons, softer missing momentum spectrum, a larger multiplicity of (softer) objects, or some combination of several of these ``escape mechanisms''.

Regardless of the precise cause of the looser bounds that apply in these models, all of them predict new physics not to show up (or at least not only) in very spectacular signatures, but in general they predict a large set of less spectacular signals, which tend to give rise to proton scattering final states that resemble very closely those of Standard Model scatterings. In view of such similarity of new physics events and Standard Model ones, it is extremely important to have under good theoretical control the predictions for Standard Model processes. Therefore the magnificent high-accuracy agreement between latest theory calculations and Run I measurements can be turned into a launch pad to attack the search of new physics with a new strategy, which leverages the precise knowledge of the Standard Model background, rather the absence of background.  

\section{A Precision Observables Program On The Top Quark}
Given that many extension of the Standard Model are built around, and motivated by, the peculiarities of the top-Higgs sector, the possibility to test with high precision the properties of the top quark at the Large Hadron Collider is very exciting. Also motivated by this perspective, the search of supersymmetric partners of the top quark (and of the bottom quark and Higgs boson) have become a standard theme of experimental investigation. Well organized summaries of the exclusion of simplified models with just a light stop and a light bino-like neutralino are routinely presented, e.g. in \cite{ATLAS-Collaboration:2015xw}. These summaries well exemplify how standard search strategies are in good position to cover most scenarios of new physics, still many scenarios remain difficult to attack. For instance for the stop-neutralino simplified model when stops become light, in the range of the top quark mass, the summary of Ref.~\cite{ATLAS-Collaboration:2015xw} shows the shortcoming of simplest searches and models with mass spectra characterized by $m_{\tilde{t}} - m_{\chi} < m_{t} $ need to investigated with specially crafted strategies, {\it e.g.} looking for mono-jet signals, soft leptons and for signatures that insist of the $W$ bosons, rather than the top quarks, to identify the decay products of the stops. 

Despite these multiple attempts to attack this, after all pretty large, fraction of the parameter space of the stop-neutralino simplified model, large portions of the light stop parameter space are still uncovered. Given how basic is the simplified model under study, and also in view of its possible role of proxy towards a large domain of ``top-like'' new physics, full-coverage of this plane  automatically emerges as a high priority task for Run II physics programs. To this end, as will be clear in a moment, precision top quark physics will be an essential asset to attain success in this endeavor. 

A latest addition to the summary on the progress in search of new physics~\cite{CMS-PAS-TOP-14-023,Aad:2014mfk}, in fact, goes precisely in the direction of using precisely computable distributions of top quark physics observables, such as the angular separation of two leptons  \cite{Han:2012ve}, to test the presence of top-like new physics with different spin. Also the measurement of the top pair total cross-section has been recently used to put bound on new physics that give rise to top-like final states and therefore affects the cross-section measurement~\cite{Czakon:2014qf,ATLAS-Collaboration:2014dp,ATLAS-Collaboration:2015xw}. For both these observables an extension of the present use can be envisaged for Run II, where a full exploration of bounds that can be obtained from these observables for generic stop mass (both below and above the top mass) and generic neutralino mass can be pursued. 

It is worth to stress that the reach of searches for new physics in precision top quark observables, {\it e.g.} the aforementioned total cross-section and spin-correlation variables, goes much beyond that of model specific searches for new physics - though sometimes less powerful, this approach always probes a large set of new physics scenarios. In fact, any scenario that modifies significantly top quark production or decay has a chance to be observed, or excluded, looking at the fine details of these quantities well under theory control.

At Run II we can  envision a new large class of observables to be used to test new physics in the top quark sample. These are the observables used for the determination of the top quark mass, which is presently extracted from Large Hadron Collider collisions to an astonishing sub-percent precision~\cite{TOPLHCWGweb,CMS-Collaboration:2015qy}. The demonstrated capability to measure this standard Model parameter to such high accuracy embodies the great theoretical control on the Standard Model theory for the relevant observables and the superb performances of the experiments in measuring these observables. The achieved control sets an excellent stage for searching new phenomena in deviations from Standard Model predictions in the distributions that are presently used to measure the top quark mass. 

Some types of new physics might induce a shift in the measured top quark mass~\cite{Eifert:2014yu}, but this is not the sole possibility. In fact the top quark mass affects certain features of the relevant distributions, while new physics in general can present itself in several features of each distribution and in many distributions at the same time. Therefore a new physics search program, which cannot be reduced to a pure reinterpretation of the mass measurement, can be envisaged for Run II. In general this way to search new physics goes beyond, although gets started from, pure Standard Model measurements and the large amount of data that Run II will be the enabling factor for this kind of searches for new physics.
In this respect the large and varied program of measurements of the top quark mass from many different quantities~\cite{CMS-PAS-TOP-15-002,Agashe:2013sw,ATLAS-Collaboration:2015qm,Alioli:2013mz,Kawabata:2014lr,Juste:2013sp,Cortiana:2015hp,Corcella:2015hb,Frixione:2014jk,CMS-PAS-TOP-11-027,CMS-PAS-TOP-12-030}, is a precious asset to test the large landscape of new physics scenarios that can manifest at the Large Hadron Collider. 

\begin{figure} \centerline{%
\includegraphics[width=.85\textwidth]{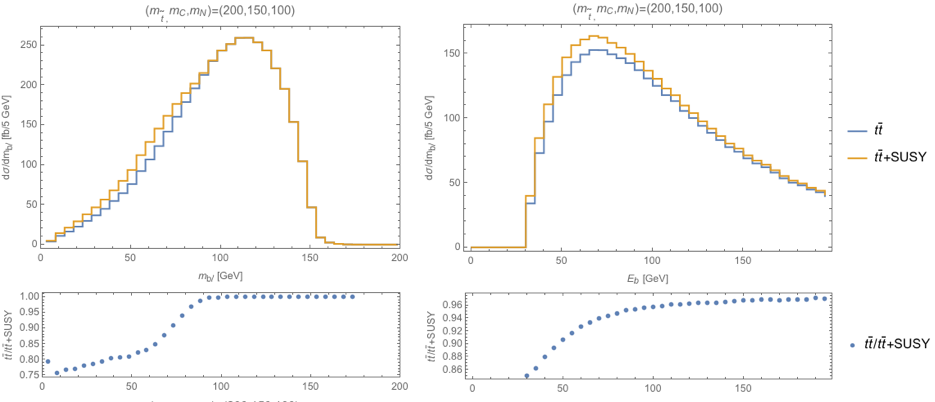}}
\caption{Possible deviation from the Standard Model shape of the lepton-b-jet invariant mass distribution and the b-jet energy distribution used for the top quark mass measurement in Ref.~\cite{CMS-PAS-TOP-14-014} and Ref.~\cite{CMS-PAS-TOP-15-002}. The bottom panel shows the ratio of the new physics over the Standard Model prediction. The effect of new physics is most apparent in certain regions of the distribution, allowing for calibration of the Standard Model prediction elsewhere in the data and test of the new physics hypothesis in the sensitive region. } \label{fig:stopcharginoneutralino}
\end{figure}

The effect of top-like new physics from supersymmetry can be very apparent in certain distribution. An example is represented for a stop-charigno-neutralino simplified model in Figure \ref{fig:stopcharginoneutralino} for the decay \begin{equation}\tilde{t} \to \tilde{\chi}^{+} b \to \ell \nu \tilde{\chi}^{0} b \end{equation}
 for a choice of masses $m_{\tilde{t}}=200\textrm{ GeV},\,m_{\tilde{\chi}^{+}}=150\textrm{ GeV},\,m_{\tilde{\chi}^{0}}=100\textrm{ GeV},\,$ on which only very mild (and uncertain) bounds exists~\cite{ATLAS-Collaboration:2014eu,ATLAS-Collaboration:2014fj,Aad:2013ija}. The new physics spectrum is characterized by mass splittings similar, but not identical, to those of the $t \to W b \to \ell \nu b$ decay of the Standard Model. The similarity of the mass scale and mass splittings renders very difficult to single out this extra source of top-like events with gross-grained techniques, such as the search for an excess in a high-pT tail of a distribution. In fact this spectrum gives rise to softer decay products (in particular to softer b quarks) than ordinary top decay, which is a background for this search. Therefore, only using very precises probes of the mass splittings, such as the observables $m_{b,\ell}$ and $E_{b}$ shown in Figure~\ref{fig:stopcharginoneutralino}, one is able to highlight the subtle difference between top decay and its supersymmetric cousin $\tilde{t} \to \tilde{\chi}^{+} b \to \ell \nu \tilde{\chi}^{0} b$.

Further opportunities for testing new physics in a broad manner thanks to precise studies of top quark properties exists in several branches of top physics studies already carried out at the Large Hadron Collider. For instance the top quark decay into $\tau$ final states $$t \to b \tau \nu$$ not mediated by a $W$ boson, has reached sensitivity to decay modes with branching ratio a fraction of that of the Standard Model \cite{CMS-PAS-HIG-12-052} and presently probes very effectively light charged Higgs boson that arise in supersymmetric models. The precise knowledge of the distributions that are used to test the light charged Higgs boson hypothesis can be used to test contact operators that mediate the same decay of the top quark, turning this search in a much broader scoped one. More in general the present status of this and other searches~\cite{CMS-Collaboration:2013vee,Davidson:2015lp}, suggests that the present knowledge of top quark decay has reached a level that enables a broad search for new physics that embraces all Standard Model final states of top decay and new ones. In this perspective we can envisage a generalization of present global determinations of the Standard Model top quark branching rations~\cite{ATLAS-Collaboration:2015hh} to encompass a broader set of observables and final states, to test more widely the presence of tiny deviations from the Standard Model in a {\it global analysis} of the properties of the top quark. The combined use of several observables, {\it e.g.} from the several final states measured, will be a point of strength of this global approach. In view of the results of Run I, this strategy will  certainly be worth pursuing at Run II both to sharpen our knowledge of the top quark in the Standard Model and at the same time test a large class of new physics scenario, in particular those that might be {\it most elusive in standard searches}.

Many more opportunities wait to be caught exploiting the large amount of Run II top quark data. A particularly useful technique can be imagined to search new physics in top-like final states by exploiting one side of the $t\bar{t}$ event as trigger, for instance requiring a leptonic top, and scrutinizing the other half of the event in search for deviations of top quark properties from the Standard Model prediction. This strategy can highlight the presence of a small branching ratio of the top quark in supersymmetric particles, {\it e.g.} in light stop and neutralino (which can easily be at the percent level \cite{Djouadi:1996fk}):
\begin{equation}t \to \tilde{t}\, \tilde{\chi}^{0} \to jets\,,\end{equation} which can arise for light stop and light neutralino in R-parity violating models with sizable $UDD$ interactions. This type of decay might be highlighted by studies of standard Model properties such as Refs.~\cite{CMS-Collaboration:2014wl,ATLAS-Collaboration:2015hh,Alwall:2007iw}, which, still being general searches for deviations from the Standard Model, can be suitably targeted towards this type of new physics, exploiting obvious features as the presence of a on-shell dijet resonance from the stop decay into jets, or the multi-jet decay of the neutralino, possibly enriched by a detectably displaced decay of the latter. On a similar note, the study of top quark decay can help to discover, or put bounds on,  supersymmetric particles that do not carry color but experience a sizable $LQD$ RPV interaction with the top quark, such as {\it staus} that can appear in the top quark decay
 \begin{equation} t \to b \tilde{\tau} \to b +jets\,.\end{equation}
\section{Conclusions}
The large production cross-section of the top quark at the Large Hadron Collider, together with the demonstrated capabilities of the experiments to measure top quark properties to high precision, and latest improvements of  high precision theory calculations allow to envision a rich program of precision physics on the top quark at Run II of the Large Hadron Collider. Certain items of this physics program can be considered standard, for instance the study of top quark couplings and the production of top quarks in more rare reactions such as production in association with other states. 

In this contribution I have argued that plenty of opportunities lays ahead if new physics is sought for in top quark properties, especially in scenarios of new physics that tend to be most elusive to the standard ``high-pT'' approach. These examples of new physics include new states with mass around that of the top quark, which have revealed severe limitations of the more generic (and more widespread)  search approaches. 
I have identified a rich program of studies of the distributions that are presently used to measure the top quark mass, as to turn these distributions in precision tests for the presence of top-like new physics. Furthermore I have highlighted opportunities in the global study of the decay of top quark in all Standard Model channels, and new ones, as suggested by the addition of contact operators involving top quarks or by explicit models such as the examples in R-parity conserving and R-parity violating supersymmetry discussed above.

\section*{Acknowledgments}
RF thanks the organizers of the LHCP 2015 conference and the LHC Physics Centre at CERN for support. Furthermore RF thanks Gabriele Ferretti, Christoffer Petersson, Giacomo Polesello, and Riccardo Torre for discussions.


\providecommand{\href}[2]{#2}\begingroup\raggedright\endgroup

\end{document}